\documentclass[twocolumn]{aastex631}
\usepackage{amsmath}
\usepackage{amssymb}
\usepackage{graphicx}
\usepackage{xcolor}
\usepackage{widetable}
\usepackage{tabularx}
\usepackage{booktabs}
\usepackage{epsf}
\usepackage{longtable}
\usepackage{mhchem}

\begin{document}

\title{A Possible Additional Formation Pathway for the Interstellar Diatomic SiS}

\author[0000-0003-4716-8225]{Ryan C. Fortenberry}
\affiliation{Department of Chemistry \& Biochemistry, University of Mississippi,\\
University, Mississippi, 38677, United States}

\author[0000-0003-1254-4817]{Brett A. McGuire}
\affiliation{Department of Chemistry, Massachusetts Institute of Technology,\\
Cambridge, Massachusetts, 02139, United States}

\date{\today}

\begin{abstract}

The formation of silicon monosulfide (SiS) in space appears to be a difficult
process, but the present work is showing that a previously excluded pathway may
contribute to its astronomical abundance.  {Reaction of the radicals SH +
SiH produces SiS with a submerged transition state and generates a stabilizing
H$_2$ molecule as a product to dissipate the kinetic energy.  Such is a textbook
chemical reaction for favorable gas-phase chemistry.  While previously proposed
mechanisms reacting atomic sulfur and silicon with SiH, SH, and H$_2$S will
still be major contributors to the production of SiS, an abundance of SiS in
certain regions could be a marker for the presence of SiH where it has
previously been unobserved.}  These quantum chemically-computed reaction
profiles imply that the silicon-chalcogen chemistry of molecular clouds, shocked
regions, or protoplanetary disks may be richer than previously thought.
{Quantum chemical spectral data for the intermediate $cis$- and
$trans$-HSiSH are also provided in order to aid in their potential spectroscopic
characterization.}

\end{abstract}


\section{Introduction} \label{sec:intro}


{
Silicon monosulfide (SiS) has been observed in a variety of astronomical
environments since it was first observed in the circumstellar environment of the evolved carbon star IRC+10216 in 1975 \citep{Morris75}.
Currently, SiS is believed to be created from the following set of reactions with their percent contributions given for SiS formation over $t=1\times10^3$ yr assuming a molecular cloud with $T=10$ K, $n(\mathrm{H}_2)=2\times10^4$ cm$^{-3}$ and a cosmic ray ionisation rate of $1.3\times10^{-17}$ s$^{-1}$ \citep{mendoza2024new}:
\begin{eqnarray}
\mathrm{Si} + \mathrm{SH} \rightarrow \mathrm{H} + \mathrm{SiS} \hspace{0.5cm} & 30.6\% \\
\mathrm{S} + \mathrm{SiC} \rightarrow \mathrm{C} + \mathrm{SiS} \hspace{0.5cm} & 26.5\% \\
\mathrm{Si} + \mathrm{SO} \rightarrow \mathrm{O} + \mathrm{SiS} \hspace{0.5cm} & 23.8\% \\
\mathrm{HSiS}^{+} + e^{-} \rightarrow \mathrm{H} + \mathrm{SiS} \hspace{0.5cm} & 18.3\% \\
\mathrm{Si} + \mathrm{H}_2\mathrm{S} \rightarrow \mathrm{H}_2 + \mathrm{SiS} \hspace{0.5cm} & 0.70\% 
\end{eqnarray}
The above reactions are built upon a large body of work from several theoretical and experimental groups \citep{Willacy:1998:676, Paiva:2018:1858, Lai01, Doddipatla:2021:eabd4044}.  The last of these above contributions to the formation of SiS  (Reaction 5) shows that even minor contributors can have a role to play in the overall chemical modeling for any potential reaction mechanism.  While other mechanisms have been suggested \citep{Rosi:2018:87, Zanchet:2018:38, Campanha:2022:369}, the above have risen to the top as the most viable mechanisms for current models that are in line with current observations of SiS \citep{mendoza2024new}.  However, other reactions could still be present as contributions to SiS abunance, and these have exciting correlations with currently hypothesized but as-of-yet undetected molecules.  
}

Recent work has shown that isolated molecule reaction pathways starting from
water and metal hydrides can lead to both simple, diatomic metal oxides as well
as larger clusters of inorganic oxides \citep{Grosselin22, Flint23SiO, Firth24};
a similar case is also present for aluminum hydride plus ammonia
\citep{Palmer24}.  {These reactions exhibit barrierless entrance intermediates, showcase submerged reaction barriers, and generate ubiquitous H$_2$ as a stabilizing leaving group}.  Such molecular behavior allows for the question as to whether
or not silicon may perform in a related fashion in the same way as its periodic
table third-row mates aluminum and magnesium will.  Additionally, the question
is also open if hydrogen sulfide or SH can replace water or the hydroxyl radical
in similar reactions.  AlO can form from AlH + OH \citep{Firth24}, {and such leads to the suggestion that
radical-radical reactions extrapolated to SiH + SH may proceed to SiS}.

Hence, the present discussion will examine the reactions of {silicon and sulfur hydrides and dihydrides} with one another in order to explore additional
means for formation of SiS beyond those reactions with atomic species:
\begin{eqnarray}
\mathrm{SiH} + \mathrm{SH} &\rightarrow \mathrm{H}_2 + \mathrm{SiS} \\
\mathrm{SiH}_2 + \mathrm{H}_2\mathrm{S} &\rightarrow 2\mathrm{H}_2 + \mathrm{SiS}.
\end{eqnarray}
Again, H$_2$S is known in astronomical regions
\citep{Thaddeus72} {as is the SH radical}
\citep{Neufeld12}.  The SiH radical has been observed in
the Sun \citep{Wohl71} implying that it could be present elsewhere
\citep{Yurchenko17}.  While clues have been given to show that SiH may be
present in the ISM \citep{Schilke:1997:293}, conclusive detection has yet to be reported \citep{2021Census}.  However, if
this molecule contributes to the formation of the known SiS molecule, renewed
attempts for detection of silicon monohydride are warranted.  SiH$_2$, however, has yet to be
observed despite searches for it \citep{Avery94}, but its inclusion is retained
herein in order to see what, if any, role it could play in the formation of SiS.  Regardless, a more {expansive} set of possibilities will be generated for
the creation of SiS.

\section{Computational Details}

All minima are optimized and real harmonic frequencies confirmed with coupled
cluster theory \citep{ccreview, Shavitt09} at the singles, doubles, and
perturbative triples level \citep{Rag89} under the F12 explicitly correlated
framework \citep{Adler07, Knizia09} with a correlation consistent triple-$\zeta$
basis set \citep{Dunning89, Peterson08} giving the CCSD(T)-F12b/cc-pVTZ-F12 (or
just F12-TZ) level of theory within the MOLPRO2022.1 quantum chemistry program
\citep{MOLPRO22, MOLPRO-WIREs}. Transition states are optimized at the
B3LYP/aug-cc-pVTZ level within Gaussian16 \citep{B3, LYP86, LYP88, g16}, but
F12-TZ single point energies are computed at these geometries to provide
consistent energetic profiles and to address any potential pitfalls in energy
B3LYP could generate.  All relative energies are computed including harmonic
zero-point energy corrections from the optimized geometries and the
corresponding level of theory.

The spectral data for the $trans$- and $cis$-HSiSH conformers are provided
herein.  These are computed from the so-called F12-TcCR \citep{Watrous21}
quartic forice field (QFF).  The QFF is a fourth-order Taylor series expansion
of the internuclear Hamiltonian and has been discussed in detail for its
relevance for astrochemical spectroscopic reference data
\citep{Fortenberry19QFF, Fortenberry22, Fortenberry24JPCA}.  The energy
surface defining the QFF is comprised of F12-TZ energies inclusive of core
electron correlation as well as canonical CCSD(T) with Douglas-Kroll scalar
relativity included.  The QFFs are computed via normal coordinates in an
automated fashion and fed through an updated second-order vibrational
perturbation theory (VPT2) code \citep{Watson77, Mills72, Papousek82} written in
\textsc{Rust} called \textsc{pbqff} \citep{Westbrook23} built upon the existing
\textsc{Spectro} code \citep{spectro91}.

\section{Results \& Discussion}

\subsection{SiS Formation}

 \begin{figure*} 
   \centering
   \includegraphics[width=\textwidth]{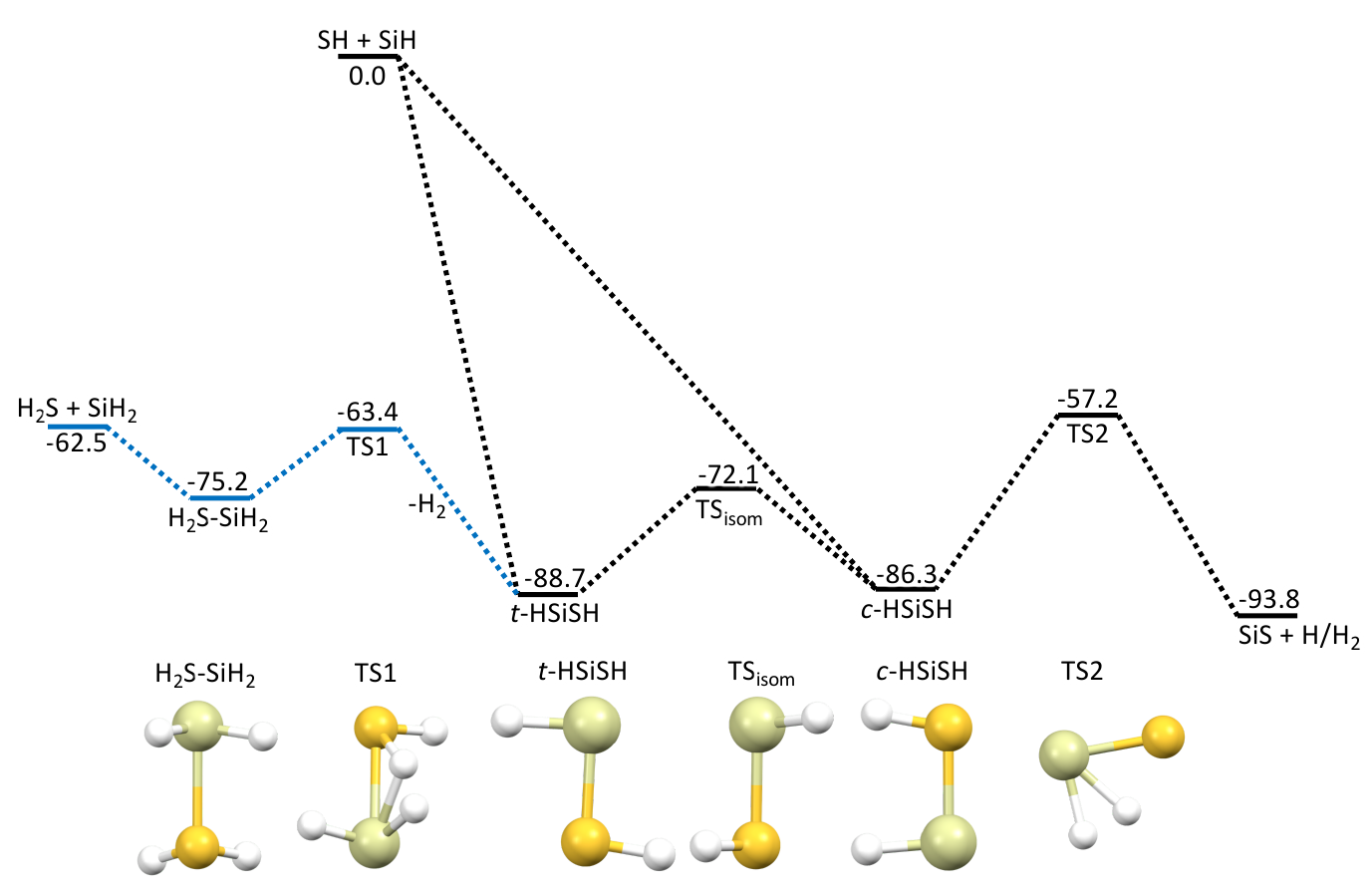}
   \caption{The F12-TZ reaction pathway for the formation of SiS {from SiHH + SH and SiH$_2$ + H$_2$S.  All energies are in kcal mol$^{-1}$. Molecular structures are given at the bottom with H=white, S=Yellow, \& Si=Orange.}}
 \label{SiS}
 \end{figure*}

{The additionally suggested formation pathways} of $^1 \Sigma^+$ SiS are shown in Fig.~\ref{SiS}. First
of all, the new reaction of SH + SiH defines the zero of energy here for these
starting materials.  The association of these two molecules can proceed to
either a $trans$- or $cis$-HSiSH intermediate.  If the lower-energy
$trans$-HSiSH (-88.7 kcal mol$^{-1}$) conformer is formed first, it can
isomerize into the $cis$ over a submerged and relatively small barrier.  Once
the two hydrogen atoms are on the same side of the molecule, they can associate
into a transition state (TS2) at -57.2 kcal mol$^{-1}$ below the starting
materials where the H$-$H bond forms.  This is nearly identical to the behavior
noted for aluminum/magnesium hydrides plus water/OH and in corroboration, in
part, with \cite{Lai01}.  At this point, the hydrogen molecule leaves producing
SiS at -93.8 kcal mol$^{-1}$ below the starting materials, and the expulsion of
the hydrogen molecule will stabilize the system as it can dissipate the excess
energy kinetically.  Such a reaction profile, especially from the SH + SiH
starting materials through $cis$-HSiSH has one barrier to overcome that is
well-submerged below the starting materials and follows a similar pattern as
that suggested for formation of the known ethynyl cyclopropenylidene
\citep{Cernicharo21c3hc2h, Fortenberry21}.  The climb from the $cis$-HSiSH
minimum intermediate to TS2 is 29.1 kcal mol$^{-1}$, but this would benefit from
the -86.3 kcal mol$^{-1}$ energy gained in the association in the first place
likely rendering such an up-hill pathway from $cis$-HSiSH to TS2 moot. 

Additionally, Fig.~\ref{SiS} also explores the reaction of H$_2$S with SiH$_2$
in spite of the lack of SiH$_2$ observation.  While this process forms a stable
intermediate (H$_2$S-SiH$_2$) resulting from a dative bond between a lone pair
on the sulfur and the empty $p$ orbital on the silicon as shown in related work
\citep{Grosselin22, Flint23SiO, Palmer24, Firth24}, the creation of TS1 where an
H$_2$ molecule is prepared for departure from the system is barely exothermic
lying -0.9 kcal mol$^{-1}$ below the pair of dihydride starting materials.  While
this would once more produce H$_2$, the effective ``solvent of the universe''
\citep{Woon23, Fortenberry24JPCA}, the barrier is likely too close to the
reactants for any population to traverse this saddle.  However, shocks, disks,
or regions with effective temperatures at or above 100 K could begin to see some
population of H$_2$ and $trans$-HSiSH form in this manner, but, again, this
would require enough SiH$_2$ to be on hand in the first place likely reducing
the role that this reaction would play.  If H$_2$S + SiH$_2$ could traverse TS1
and form $trans$-HSiSH, it would initially kick off an H$_2$ in the process
which would dissipate the kinetic energy and stabilize HSiSH.  However, TS2 is
higher than TS1 implying that at best this reaction would form $trans$-HSiSH and
maybe would isomerize to $cis$-HSiSH.  It would be highly unlikely to contribute
to any formation of SiS.

Finally, the reaction of SiH + H$_2$S $\rightarrow$ SiSH + H$_2$ $\rightarrow$
SiS + H$_2$ + H has also been computed in this work but is not shown in
Fig.~\ref{SiS}.  The reason is that while this reaction is net exothermic, the
first TS for creation of H$_2$ in the first step has an energy of -0.8 kcal
mol$^{-1}$ relative to the reactants.  Within the accuracies of the approach,
this is too close to say that it would confidently be a submerged barrier.
Additionally, the breaking of the S$-$H bond to form the products puts the final
energy at only -3.7 kcal/mol, again, not low enough for this to be confidently
exothermic.  Even if exothermic, such a small value would make this reaction
relatively slow and would likely be a minor contributor to formation of SiS at
best.

\subsection{Spectral Characterization of $trans$- and $cis$-HSiSH}

\begin{centering}
\begin{table}
    \caption{Harmonic ($\omega$) and Anharmonic ($\nu$) Vibrational Frequencies (in cm$^{-1}$), Intensities ($f$ in parentheses; in km/mol)$^a$, Dipole Moments (D), and Rotational Spectroscopic Data (in cm$^{-1}$) for $trans$- and $cis$-HSiSH from the F12-TcCR QFF.}
    \label{Spec}
    \begin{tabular}{l|ll}
     & $trans$-HSiSH & $cis$-HSiSH \\
     \hline
     $\omega_1$ (a') & 2700.1 (1)  & 2713.6 (1)  \\
     $\omega_2$ (a') & 2070.1 (232)& 2066.8 (245)\\
     $\omega_3$ (a') &  920.6 (35) &  813.7 (54) \\
     $\omega_4$ (a') &  637.1 (17) &  666.2 (8)  \\
     $\omega_5$ (a'')&  633.1 (1)  &  541.9 (15) \\
     $\omega_6$ (a') &  526.4 (43) &  516.6 (51) \\
     \hline
     $\nu_1$ (a') & 2586.4 (1)  & 2600.2 (1)  \\
     $\nu_2$ (a') & 1982.1 (234)& 1987.8 (249)\\
     $\nu_3$ (a') &  896.9 (34) &  794.1 (50) \\
     $\nu_4$ (a') &  606.5 (17) &  651.0 (7)  \\
     $\nu_5$ (a'')&  615.1 (1)  &  508.8 (15) \\
     $\nu_6$ (a') &  511.8 (43) &  499.5 (51) \\
     Zero-point   & 3691.5      & 3610.5 \\
     \hline
     $A_e$ & 4.293204 & 4.322487 \\
     $B_e$ & 0.243442 & 0.239242 \\
     $C_e$ & 0.230379 & 0.226694 \\
     $A_0$ & 4.259478 & 4.299109 \\
     $B_0$ & 0.242050 & 0.238045 \\
     $C_0$ & 0.228762 & 0.225278 \\
     $\Delta_J$ ($\times 10^{-6}$) & 0.186 & 0.182 \\
     $\Delta_K$ ($\times 10^{-6}$) & 65.255 & 69.300 \\
     $\Delta_{JK}$ ($\times 10^{-6}$) & 2.125 & 3.065 \\
     $\delta_J$ ($\times 10^{-6}$) & 0.010 & 0.009 \\
     $\delta_K$ ($\times 10^{-6}$) & 1.464 & 1.783 \\
     $\Phi_J$ ($\times 10^{-14}$) & -6.772 & -9.574 \\
     $\Phi_K$ ($\times 10^{-7}$) & -3.171 & -1.141 \\
     $\Phi_{JK}$ ($\times 10^{-7}$) & -1.579 & -0.761 \\
     $\Phi_{KJ}$ ($\times 10^{-7}$) & 5.318 & 2.588 \\
     $\phi_J$ ($\times 10^{-14}$) & 15.576 & 13.870 \\
     $\phi_{JK}$ ($\times 10^{-7}$) & 17.405 & 12.703 \\
     $\phi_K$ ($\times 10^{-7}$) & 479.536 & 390.488 \\
     $\mu$ & 0.59 & 1.04 \\
       \hline 
    \end{tabular}
\\$^a$Computed at the B3LYP/aug-cc-pVTZ level.
\end{table}
\end{centering}

{While the lifetimes of the HSiSH intermediates will likely be too short for astrophysical observation stemming from the reaction of SiH + SH, they are the products of SiH$_2$ + H$_2$S.  Hence, the presence of either form of HSiSH could indicate some abundance of SiH$_2$.  Even so, high-resolution laboratory experiments could be able to observe $cis$- or $trans$-HSiSH during its brief epochs of existence in either reaction.}

As such and in order to aid in potential observation or laboratory characterization of any
reaction pathway leading to SiS as described in Fig.~\ref{SiS}, detectable
spectral features must be provided for the molecular species involved.
Notably, this includes $trans$- and $cis$-HSiSH (Table \ref{Spec}) which have
not been analyzed in significant detail, yet, in the literature unlike their
H$_2$SiS isomer \citep{McCarthy11}.  Both conformers are rotationally active,
but their dipole moments are not so large as to indicate that they would
immediately be targets of radioastronomical searches. The $trans$ conformer
dipole moment of 0.59 D is lower than the $cis$ at 1.04 D implying that the
higher-energy $cis$ conformer would be more observable akin to recent observations
of carbonic acid \citep{Sanz23}.  Neither conformer has any contribution to the
dipole moment from the coordinate largely defined from the vector of the Si$-$S
bond; all of the dipole moment vector effectively is perpendicular to it.  Both conformers have very
similar rotational constants owing to the fact that both are dominated by the
much more massive Si and S atoms regardless of the orientation of the much
smaller hydrogen atoms.  Case in point, both conformers are near prolate with
$k=-0.99$ for each. The full set of principal ($A_0$, $B_0$, \& $C_0$), quartic ($\Delta$/$\delta$), and sextic ($\Phi$/$\phi$) rotational constants are given in Table \ref{Spec}.

This similarity is continued in the fundamental vibrational frequencies also
given in Table \Ref{Spec}. The S-H and Si-H stretches ($\nu_1$ \& $\nu_2$,
respectively) are within 15 cm$^{-1}$ or less of one another between isomers,
and $\nu_4$ \& $\nu_6$ differ between isomers by a little more.  However,
$\nu_3$ \& $\nu_5$ are more than 100 cm$^{-1}$ apart between isomers.  These
last two modes correspond to the ``symmetric'' bend as well as the out-of-plane
bend, respectively.  While not totally ``symmetric,'' the normal mode
coordinates for $\nu_3$ show both hydrogen-terminated bond angles changing
simultaneously. Even though this is a free motion in the $trans$-HSiSH isomer, the two
hydrogen atoms hinder one another significantly in the $cis$ form explaining the
reduction in frequency therein. This is also present in the $\nu_5$ $a''$
frequency where the $cis$ conformer is more hindered in its rotation.

With the launch of \textit{JWST}, the IR features of such molecules may become
notable for observation, the $\nu_2$ Si-H stretch most notably for these
conformers of HSiSH.  This intensity of $\sim 240$ km mol$^{-1}$ is more than
three times the antisymmetric stretch in water giving it a large transition
dipole {making it potentially observeable even if interstellar lifetimes of HSiSH in either conformer are relatively short}.  These frequencies correspond to $\sim5.05$ $\mu$m right at the edge of
what JWST's NIRSpec instrument can observe, but first-look spectra indicate
features in this region and slightly beyond \citep{Boersma23} that are present
in JWST observations.  In using IR spectra to distinguish conformers most likely to be created
in the laboratory, the $\nu_3$ frequencies would be the best option, again, due
to their separation, but also because they have notable intensities, in line with
what would be expected for IR features of organic molecules.

\subsection{Astrochemical Implications}

While SiS is not a rare interstellar species, it is certainly not widely observed and has, thus far, primarily been identified in evolved stars (e.g. IRC+10216; \citealt{Gong:2017:54}).  This diatomic molecule is seen in a small selection of other locations, including massive protostars (e.g. Orion Src I; \citealt{Wright:2020:155}), shock-associated molecular outflows in a few star-forming regions and protostellar sources (e.g. L1157-B1; \citealt{Podio:2017:L16}), and very recently in protoplanetary disks, potentially surrounding protoplanets \citep{Law:2023:L19}.  SiS emission appears to be associated with shocked regions in these sources.  Yet, as pointed out by \citet{Podio:2017:L16}, the spatial distribution of SiS differs from that of SiO, suggesting different formation pathways.  SiO is associated with the strongest shocked regions, while SiS is found offset from these locations.  

This may suggest that SiS is formed in a delayed fashion from material not directly generated by the shock, but rather from a second generation of products formed from these.  One such second-generation product is SiH, which according to \citet{Schilke:1997:293} can form from the successive dehydrogenation of silane (\ce{SiH4}):
\begin{equation}
    \ce{SiH4 + H -> SiH3 + H2}\label{sih4_h_sih3_h2}
\end{equation}
\begin{equation}
    \ce{SiH3 + H -> SiH2 + H2}
\end{equation} 
\begin{equation}
    \ce{SiH2 + H -> SiH + H2}.
\end{equation}
Reaction~\ref{sih4_h_sih3_h2} is endothermic by $\sim$1400\,K, meaning that this can likely only proceed in the very high temperatures of the post-shocked gas.  Such high-temperature chemistry, resulting in chemical abundance peaks delayed in time from the shock event itself, has been shown to be viable in chemical shock models \citep{Burkhardt:2019:32}.  

As noted in the introduction, however, the definitive presence of SiH in the interstellar medium or circumstellar environments has yet to be confirmed \citep{2021Census}.  A tentative detection of SiH was made toward Orion-KL by \citet{Schilke:2001:281}, but further searches for the molecule by \citet{Siebert:2020:22} failed to identify it using SOFIA data.  Given the work presented here that shows SiS can be efficiently formed from SiH via Reaction~\ref{sih_sh_sis_h2}
\begin{equation}
    \ce{SiH + SH -> SiS + H2}, \label{sih_sh_sis_h2}
\end{equation}
this suggests a straightforward set of observational tests that can be performed to assess the importance of this pathway.  If Reaction~\ref{sih_sh_sis_h2} is indeed dominant, then it is reasonable to expect we should be able to detect the reactants, SH and SiH, co-located with the SiS emission.  Because this is likely to be in relatively compact spatial regions concentrated around shocked gas, interferometric observations are likely mandatory.  The molecular outflow in Orion Src I is perhaps best-suited to this task.  It is well-observed, and the spatial structure of SiS is well-constrained \citep{Wright:2020:155}. 

\section{Conclusions}

{An additional pathway to SiS through SH + SiH (Reaction 6) is shown here to have many advantages for gas-phase chemistry: a large exothermicity, submerged barriers, and the formation of H$_2$ as a leaving group.  While Reactions 1 through 5 certainly cannot be discounted for their roles in creation of SiS, this additional pathway may help to further explain SiS abundance in observed regions augmenting recent work \citep{mendoza2024new}.  
Reaction 6 (SiH + SH) is dependent upon a high enough population of SiH to be present in regions where SH is also found, but such coupling between SiS abundance and the presence of SiH would give evidence for the existence of the currently undetected SiH diatomic.  Hence, if higher abundance of SiS is present, such could imply the presence of SiH if the proposed, additonal mechanism for SiS is a contributor.  Orion Src I would be a logical target for further examination of the presence of SiH. Regardless, the present work is showing that this additional mechanism (Reaction 6) should be included in reaction networks where SiS is produced.}

H$_2$S + SiH$_2$ likely will have little-to-no role in the formation of SiS. H$_2$S + SiH will have even less.  Warmer or denser regions could
alter this, and the former could lead to HSiSH formation making this a possible
molecule for future astronomical detection.  Even so, H$_2$S + SiH$_2$ still
requires the presence of the elusive SiH$_2$.  SiH has been difficult to observe
save for the closest of stars. SiH$_2$ would presumably be moreso. IR and
rotational spectral data are also provided for the HSiSH intermediate conformers
in order to aid in possible experimental or observational characterization of
these molecules and any reaction pathways in which they may participate.
Microwave observation would likely favor the $cis$ conformer, but both
conformers exhibit a large intensity for the Si-H stretch in the 5 $\mu$m range.

\section{Acknowledgments}

This work is supported by NASA Grant NNH22ZHA004C and by the University of
Mississippi's College of Liberal Arts. The Mississippi Center for
Supercomputing Research provided the computational resources for this work and
is funded in part by NSF Grant OIA-1757220.  RCF would also like to acknowledge Dr.~Vincent J.~Esposito of the NASA Ames Research Center for useful discussions related to this work.  The National Radio Astronomy Observatory is a facility of the National Science Foundation operated under cooperative agreement by Associated Universities, Inc.
 

\end{document}